\documentstyle [12pt]{article}
\begin{document}
\setcounter{page}1
\setcounter{footnote}0

\begin{flushright}
\large\bf
KIPT E96-3
\end{flushright}

\begin{center}
{\large \bf
National Science Center\\
"Kharkov Institute of Physics and Technology"\\
\vspace{5cm}
M.I.Ayzatsky\footnote{ M.I.Ayzatsky (N.I.Aizatsky)\\
National Science Center
"Kharkov Institute of Physics and Technology"\\
Akademicheskaya 1,
Kharkov, 310108, Ukraine\\
e-mail:aizatsky@nik.kharkov.ua}\\
\vspace{2cm}
ON THE PROBLEM OF THE COUPLED CAVITY CHAIN CHARACTERISTIC CALCULATIONS \\
\vspace{1.5cm}
E-Preprint\\
\vspace{6cm}
Kharkov --- 1996}
\end{center}
\newpage
\begin{abstract}

This paper presents the results of studies in elaboration of a mathematical
model of the cavity-chain slow wave structures. Considered is the problem of
coupling  of an infinitely long cylindrical cavity chain  coupled through
centerholes   in the dividing walls of finite thickness without the assumption about
smallness of any parameters. On the basis of a rigorous electrodynamic
approach it is shown that the cavity chain can be described by a set of
equations that describe the coupling of an infinite number of similar-type
modes of short-circuit cavities. At the same time the coupling coefficients
are determined by solving the basic sets of linear algebraic equations
which describe the coupling the tangential electric fields on the right
and left cylindrical hole cross-sections of the disks. The results are
given of numerical
simulations of the dependence of electrodynamic characteristics on the
number of connections between the modes taken into account. Based on these
calculations for different cavity dimensions, the number of coupling
connections that give the sufficient accuracy of the description
was determined.
\end{abstract}

\section{Introduction}

The chain of coupled cavities are widely used in the RF-engineering.
Slow wave structures on their base are most common in the accelerator
technology, finding various applications as well in RF-devices designed
to generate and amplify electromagnetic waves (see, for example
\cite{r1}-\cite{r3}). Since resonant properties of each  cavity can be
described by equations that resemble in their outward appearance the classic
equations of the resonant circuits, then, a coupled cavity chain should be
described
similarly. Such an approach to the study of properties of coupled
cavity chain (a method of equivalent circuits)
is very useful to model rf cavities. Its advantage over purely electrodynamic
methods is in its explicitness and a relative simplicity of the mathematical
analysis which is of supreme importance for the stage of primary
electrodynamic properties study and conceptual design of the system.
It is especially manifest in the development of complex structures:
a chain of cavities coupled through slots \cite{r4},
a biperodic or compensated structure \cite{r5,r6}, a detuned structure
\cite{r7,r8} and others.
However, justification of such models must be made on the base
of the electrodynamic approach which simultaneously gives their
accuracy.

The main question in utilization of the circuit model for description
of a coupled cavity chain is the possibility of truncation the number of
circuits under consideration and their connections, since a precise account
of all these factors would do more than simply eliminate the advantages of
this approach: it would make this problem mathematically unresolvable.

By doing such truncation we have to take into account the following
circumstance.
Since commonly the analysis of characteristics has to be made within a
confined frequency range, then the total initial set of circuits
can be broken down into two classes: the resonant one, representing modes
with eigen-frequencies which values are close to the frequency range under
the consideration, and the non-resonant one. If the couplings of the resonant
circuits  form the main frequency properties of the system, then the presence
of the non-resonant circuits form the properties of such couplings.
Although the amplitude of each non-resonant mode is small , their total
effect on resonant circuit couplings is considerable.

Based on the rigorous electrodynamic approach, in two cavity coupling
problem we could separate the above circuit types and also bring down the
study of influence of the non-resonant modes to the field coupling on the
boundaries dividing the cavities (\cite{r9}-\cite{r11}). This approach
allowed to preserve the explicitness of the model as a system of coupled
resonant circuits and calculate accurately the necessary coupling
coefficients.

In this work this method is used to describe an infinitely long chain
of cylindrical cavities coupled through central holes in sidewalls.
The focus of attention is paid to the calculation of the value of coupling
of cavities which have no immediate contact. The number of couplings to be
taken into account determines not only the slow-wave structure properties,
but also a possibility of their matching and tuning. The latter is of added
importance for development of inhomogeneous structures. As distinct from
the previous papers (\cite{r3}, \cite{r5,r6}, \cite{r12}), we have managed
to elaborate on a model, allowing consecutively to take into consideration
any number of couplings.

\section{Problem Definition}

Let us consider an infinite chain of similar ideally conducting co-axial
cylindrical cavities (disk-loaded wavequide) coupled through cylindrical
holes with the radius $a$  in the dividing walls of the thickness $t$.
The radii and lengths of the cavities we denote by $b$ ¨ $d$. The disks of
the $i$-th cavity we denote by the indexes $i$ ¨ $i-1$. In order to construct
a mathematical system model under consideration we will use a method of
partial cross-over regions \cite{r13}. As a main set of regions,
we take the cylindrical cavity volumes; as an auxiliary one ---
the cylinders that are co-axial with coupling holes and of the same
radii\footnote{All auxiliary region-related values will marked by the prime,
for instance, $b^{\prime}$, $\omega_{n^\prime,s^\prime}^{\prime}$}
$b^{\prime}=a$. Each of these cylinders projects into the volumes of
the adjacent cavities over the length  $d_{*}$. The lengths of these
cavities we denote by $d^{\prime}$ ($d^{\prime}=2 d_{*}+t$).
In each region of the two sets we expand
the electromagnetic field with the short-circuit resonant cavity
modes\footnote{If we employ such expansion
we must observe the fact that this expansion is complete for total electric
field within the cavity, but it is incapable of representing tangential
electric field on the hole, so that is the reason why the condition
$d_{\ast} < d$ should be imposed on $d_{\ast}$, at least, during derivation
of the set of equations (see below)}.
Below, we will consider only axially-symmetric  $…$-fields, that is why
the irrotational mode amplitudes in the appropriate expansions will be zero.
In this case for the main regions ($i$ is the cavity number) the expansions
will be as follows:
\begin{eqnarray}
\vec{E^{i}}=\sum_{n,s}e_{n,s}^{(i)}\vec{\cal E}_{n,s}^{(i)} \label{qw1}\\
\vec{H^{i}}=\sum_{n,s}h_{n,s}^{(i)}\vec{\cal H}_{n,s}^{(i)} \label{qw2}
\end{eqnarray}
where
\begin{eqnarray}
{\cal E}_{n,s,\rm z}^{(i)}=\frac{\lambda_s^2 c}{b^2 \omega_{n,s}
N_{n,s}} \cos[k_n(z-D\times i)] J_0\left(\lambda_s r/b\right),
\label{qw3} \\
{\cal E}_{n,s,\rm r}^{(i)}=\frac{\lambda_s k_n c}{b
\omega_{n,s} N_{n,s}}\sin[k_n(z-D\times i)] J_1\left(\lambda_s
r/b\right), \label{qw4} \\
{\cal H}_{n,s,\rm \phi}^{(i)}=-i \frac{\lambda_s }{b
N_{n,s}}\cos[k_n(z-D\times i)] J_1\left(\lambda_s
r/b \right), \label{qw5}
\end{eqnarray}
$$
s=1,2,3 \ldots \infty; \ n=0,1,2 \ldots \infty;
J_0\left(\lambda_s\right)=0;
$$
$$
\omega_{n,s}=c\sqrt{k_n^{2}+\lambda_s^2/b^2}; \ k_{n}=\pi n/d;
\ N_{n,s}=\sqrt{\pi \epsilon_n d \lambda_s^2
J_1^2(\lambda_s)/2};
$$
$$
\epsilon_n=\left\{
\begin{array}{lr}
2,&n=0,\\
1,&n\not= 0,
\end{array}
\right. \
$$
The mode set (\ref{qw3}-\ref{qw5}) satisfies the following orthonormality
conditions :
\begin{equation}
\int_v\vec{\cal E}_{n,s}^{(i)}\vec{\cal E}_{n^\prime,s^\prime}^{(i)\ast}dV=
\int_v\vec{\cal H}_{n,s}^{(i)}\vec{\cal H}_{n^\prime,s^\prime}^{(i)\ast}dV=
\delta_{n,n^\prime}\delta_{s,s^\prime}. \label{qw6}
\end{equation}

For the auxiliary regions the field expansion takes on the form similar to
(\ref{qw1},\ref{qw2}), with the eigen-functions being derived from the
formulae (\ref{qw3}--\ref{qw5}) by the way of substitution
$b\rightarrow a, d\rightarrow d^{\prime}$.

The coefficients in the expansion (\ref{qw1}) are determined by the tangential
components of electric field on the boundaries of regions of choice
\begin{equation}
\left(\omega_{n,s}^{(i)2}-\omega^2\right) e_{n,s}^{(i)}=-ic\omega_{n,s}^{(i)}
\int_{S}\left[\vec{E}\vec{\cal H}_{n,s}^{\ast(i)}\right]d\vec{s}. \label{qw7}
\end{equation}

Since the electric field tangential component on a metallic surface is zero,
then, in Eq.(\ref{qw7}) the integration surfaces for the main regions will
be circles located on the opposite cavity walls, while for the auxiliary ones
--- two cylindrical surfaces and two circles, over which these regions are in
contact with the former two regions. Remembering this, we derive from
(\ref{qw7}) the following:
\begin{equation}
\left(\omega_{k,l}^{2}-\omega^2\right) e_{k,l}^{(i)}=
\sum_{n^{\prime},s^{\prime}}\left( e_{n^{\prime},s^{\prime}}^{\prime\,(i)}
L_{n^{\prime},s^{\prime},k,l}^{(1)}+
e_{n^{\prime},s^{\prime}}^{\prime\,(i-1)}
L_{n^{\prime},s^{\prime},k,l}^{(2)}\right),
 \label{qw8}
\end{equation}
\begin{equation}
e_{n^{\prime},s^{\prime}}^{\prime\,(i)}=\sum_{n,s}
\left(
e_{n,s}^{(i)} T_{n,s,n^{\prime},s^{\prime}}^{(1)}+
e_{n,s}^{(i+1)} T_{n,s,n^{\prime},s^{\prime}}^{(2)}
\right), \label{qw9}
\end{equation}
where
$$
L_{n^{\prime},s^{\prime},k,l}^{(1)}=-ic\omega_{k,l} 2\pi
\int_{0}^{a} rdr{\left({\cal E}_{n^{\prime},s^{\prime},\rm
r}^{\prime\,(i)} {\cal H}_{k,l,\rm{\phi}}^{(i)\,\ast}\right)}_{z=d+D\times i},
$$
$$
L_{n^{\prime},s^{\prime},k,l}^{(2)}=ic\omega_{k,l} 2\pi
\int_{0}^{a} rdr{\left({\cal E}_{n^{\prime},s^{\prime},\rm
r}^{\prime\,(i-1)} {\cal H}_{k,l,\rm{\phi}}^{(i)\,\ast}\right)}_{z=D\times i},
$$
$$
T_{n,s,n^{\prime},s^{\prime}}^{(1)}=
\frac{2\pi i c \omega_{n^{\prime},s^{\prime}}^{\prime}}
{\omega_{n^{\prime},s^{\prime}}^{\prime\,2}-\omega^2} \times
$$
$$
\times \left[
-a\int_{d-d_{\ast}+D\times i}^{d+D\times i} dz
{
\left( {\cal E}_{n,s,\rm z}^{(i)}
{\cal H}_{n^{\prime},s^{\prime},\rm{\phi}}^{\prime\,(i)\,\ast}\right)
}_{r=a}- \\
\int_0^a rdr
{
\left({\cal E}_{n,s,\rm r}^{(i)}
{\cal H}_{n^{\prime},s^{\prime},\rm{\phi}}^{\prime\,(i)\,\ast}\right)
}_{z=d-d_{\ast}+D\times i}
\right],
$$
$$
T_{n,s,n^{\prime},s^{\prime}}^{(2)}=
\frac{2\pi i c \omega_{n^{\prime},s^{\prime}}^{\prime}}
{\omega_{n^{\prime},s^{\prime}}^{\prime\,2}-\omega^2} \times
$$
$$
\times \left[
-a\int_{D+D\times i}^{D+d_{\ast}+D\times i} dz
{
\left( {\cal E}_{n,s,\rm z}^{(i+1)}
{\cal H}_{n^{\prime},s^{\prime},\rm{\phi}}^{\prime\,(i)\,\ast}\right)
}_{r=a}- \\
\int_0^a rdr
{
\left({\cal E}_{n,s,\rm r}^{(i+1)}
{\cal H}_{n^{\prime},s^{\prime},\rm{\phi}}^{\prime\,(i)\,\ast}\right)
}_{z=D+d_{\ast}+D\times i}
\right].
$$

By substituting (\ref{qw9}) into (\ref{qw8}), and also by introducing new
variables instead of $e_{k,\ell}^{(i)}$ for convenience
$$
a_{k,l}^{(i)}=e_{k,l}^{(i)} \frac{\lambda_l J_0 \left(\lambda_l a/b\right)}
{\omega_{k,l} \sqrt{\epsilon_k} J_1 (\lambda_l)},
$$
we derive a set of equations for mode amplitudes only in the main regions:
$$
\epsilon_k Z_{k,l} a_{k,l}^{(i)}=
\sum_{n,s}
a_{n,s}^{(i)}
\left( V_{n,s,k,l}^{(1,1)}+
V_{n,s,k,l}^{(2,2)} \right)+
$$
\begin{equation}
+\sum_{n,s}
\left(
a_{n,s}^{(i+1)} V_{n,s,k,l}^{(1,2)}+
a_{n,s}^{(i-1)} V_{n,s,k,l}^{(2,1)}
\right)
\label{qw10}
\end{equation}
where $Z_{k,l}=\omega_{k,l}^{2}-\omega^2$,
$$
V_{n,s,k,l}^{(i,j)}=(-1)^{1+i\times(1+k)+j\times(1+n)}
\alpha \gamma_{l} \times
$$
$$
\times
\sum_{s^{\prime}} \sigma_{s^{\prime},l}\Delta_{s^{\prime},n}
\left[f_{s^{\prime}}^{(i,j)}- \beta Z_{n,s}
\sigma_{s^{\prime},s} B_{n} F_{s^{\prime}}^{(i,j)}\right],
$$

$$
\alpha=4a^3 c^2 / \left(b^4 d\right),
\gamma_{l}=\lambda_l^2 J_0^2\left(\lambda_l a/b \right)/J_1(\lambda_l),
$$
$$
\sigma_{s,l}={\left( \lambda_s^2-a^2\lambda_l^2/b^2\right) }^{-1},
\Delta_{s,n}={\left[ \lambda_s^2- \Omega^2+{\left(\pi a n/d \right) }^2
\right] }^{-1},
$$
$$
\beta=2a^3/\left( c^2d \right), B_{n}=\pi n \sin(\pi n d_{\ast}/d),
$$
$$
F_s^{(i,j)}=\frac{1}{\sinh(q_s)}
\left\{
\begin{array}{lr}
\sinh[ q_s \left( 1-d_{\ast}/d^{\prime} \right) ],& i=j,\\
\sinh( q_s d_{\ast}/d^{\prime} ),& i\not=j,
\end{array}
\right.
$$
$$
f_s^{(i,j)}=\frac{\mu_s}{\sinh(q_s)}
\left\{
\begin{array}{lr}
\cosh(q_s)-\cosh[ q_s \left( 1-2 d_{\ast}/d^{\prime} \right) ],& i=j,\\
\cosh(q_s 2 d_{\ast}/d^{\prime})-1 ,& i\not=j,
\end{array}
\right.
$$
$$
q_s=\mu_s d^{\prime}/a, \; \mu_s=\sqrt{\lambda_s^2-\Omega^2},\;
\Omega=\omega a/c.
$$

The homogeneous set of Eqs.(\ref{qw10}) describes the coupling of
infinite set of resonant circuits which are the short-circuit resonant cavity
modes, and, in principle, it can be used to calculate the necessary
electrodynamic characteristics of a cavity chain. However, the set
(\ref{qw10}) has a few drawbacks that make it difficult to carry out both
analytical investigations and  numerical calculations.
Firstly, the structure of this set of equations does not yield a
possibility to obtain analytical results, in particular,  it is difficult
to take account the non-resonant modes. Secondly, this set is
two-dimensional, and it is necessary to have great calculative resources
to solve it. Our studies show that the set of Eqs.(\ref{qw10})  can be reduced
to such a form that would permit analytical studies to be made and simplify
the numerical calculations. Below we present the results of our studies.

\section{Derivation of the main set of equations}

We shall seek the amplitudes $a_{k,l}^{(i)}$, with the exception of the
fundamental modes ($(k,l)\not=(0,1)$), as:
\begin{equation}
\epsilon_k Z_{k,l} a_{k,l}^{(i)}=
\sum_{m}
a_{0,1}^{(m)} x_{k,l}^{(i,m)}. \label{qw11}
\end{equation}
Introducing instead of one sequence of unknowns
$\left\{a_{k,l}^{(i)}\right\}$, \linebreak ($i=-\infty,~+\infty$)
an infinite number of new ones  $\left\{x_{k,l}^{(i,m)}\right\}$,
($i=-\infty,+\infty$, $m=-\infty,+\infty$), we can impose $(\infty-1)$
additional conditions on the new sequences. We shall consider that
$\left\{x_{k,l}^{(i,m)}\right\}$ satisfy the equations
$$
x_{k,l}^{(i,m)}-{\sum_{n,s}}^{\prime}
\frac{1}{\epsilon_{n} Z_{n,s}}
\left[
x_{n,s}^{(i,m)}
\left(
V_{n,s,k,l}^{(1,1)}+
V_{n,s,k,l}^{(2,2)}
\right)+
\right.
$$
$$
\left.
+x_{n,s}^{(i+1,m)} V_{n,s,k,l}^{(1,2)}+
x_{n,s}^{(i-1,m)} V_{n,s,k,l}^{(2,1)}
\right]=
$$
\begin{equation}
=\left(
V_{0,1,k,l}^{(1,1)}+V_{0,1,k,l}^{(2,2)}
\right)\delta_{m,i}+V_{0,1,k,l}^{(1,2)}\delta_{m,i+1}+
V_{0,1,k,l}^{(2,1)}\delta_{m,i-1},
\label{qw12}
\end{equation}
where $(k,l)\not=(0,1)$.
In Eqs.(\ref{qw12}) and elsewhere below the prime in sums indicate
that $(n,s)\not=(0,1)$. In so doing, from (\ref{qw10}) it follows that the
amplitudes of the fundamental modes ($(k,l)=(0,1)$) should satisfy the
equations
$$
2 Z_{0,1} a_{0,1}^{(i)}=a_{0,1}^{(i)}
\left(V_{0,1,0,1}^{(1,1)}+V_{0,1,0,1}^{(2,2)} \right)+
a_{0,1}^{(i+1)} V_{0,1,0,1}^{(1,2)}+
a_{0,1}^{(i-1)} V_{0,1,0,1}^{(2,1)}+
$$
$$
+\sum_{m} a_{0,1}^{(m)}
{\sum_{n,s}}^{\prime}
\frac{1}{\epsilon_{n} Z_{n,s}}
\left[
x_{n,s}^{(i,m)}
\left(
V_{n,s,0,1}^{(1,1)}+
V_{n,s,0,1}^{(2,2)}
\right)+
\right.
$$
\begin{equation}
\left.
+x_{n,s}^{(i+1,m)} V_{n,s,0,1}^{(1,2)}+
x_{n,s}^{(i-1,m)} V_{n,s,0,1}^{(2,1)}
\right].      \label{qw13}
\end{equation}
Let's denote
\begin{equation}
v_{+,s}^{(i,m)}=f_{s}^{(1,1)} p_{+,s}^{(i+1,m)}-
f_{s}^{(1,2)} p_{-,s}^{(i,m)}-
F_{s}^{(1,1)} q_{+,s}^{(i+1,m)}+
F_{s}^{(1,2)} q_{-,s}^{(i,m)}, \label{qw14}
\end{equation}
\begin{equation}
v_{-,s}^{(i,m)}=f_{s}^{(1,1)} p_{-,s}^{(i,m)}-
f_{s}^{(1,2)} p_{+,s}^{(i+1,m)}-
F_{s}^{(1,1)} q_{-,s}^{(i,m)}+
F_{s}^{(1,2)} q_{+,s}^{(i+1,m)},   \label{qw15}
\end{equation}
where
$$
{\sum_{n,s}}^\prime
\frac{x_{n,s}^{(i,m)}}
{\epsilon_{n} Z_{n,s}}
\Delta_{s^{\prime},n}=q_{+,s^{\prime}}^{(i,m)}-\delta_{i,m}
\Delta_{s^{\prime},0}
$$
$$
{\sum_{n,s}}^\prime
(-1)^{n}\frac{x_{n,s}^{(i,m)}}
{\epsilon_{n} Z_{n,s}}
\Delta_{s^{\prime},n}=q_{-,s^{\prime}}^{(i,m)}-\delta_{i,m}
\Delta_{s^{\prime},0}
$$
$$
{\sum_{n,s}}^\prime
\frac{x_{n,s}^{(i,m)}}
{\epsilon_{n}}
\Delta_{s^{\prime},n}\sigma_{s^{\prime},s} B_{n}
=p_{+,s^{\prime}}^{(i,m)}
$$
$$
{\sum_{n,s}}^\prime
(-1)^{n}\frac{x_{n,s}^{(i,m)}}
{\epsilon_{n}}
\Delta_{s^{\prime},n}\sigma_{s^{\prime},s} B_{n}
=p_{-,s^{\prime}}^{(i,m)}
$$
Then, the set of equations (\ref{qw13}) can be written down as:
\begin{equation}
2 Z_{0,1} a_{0,1}^{(i)}=-\alpha \gamma_{1}
\sum_{m}a_{0,1}^{(m)} \sum_{s^{\prime}} \sigma_{s^{\prime},1}
\left[
v_{+,s^{\prime}}^{(i-1,m)}+v_{-,s^{\prime}}^{(i,m)}
\right], \label{qw16}
\end{equation}
As follows from (\ref{qw12}), the coefficients $v_{+,s}^{(i,m)}$ and
$v_{-,s}^{(i,m)}$  satisfy the equations
$$
v_{+,n}^{(i,m)}+\sum_{s} \left[
v_{+,s}^{(i,m)}G_{n,s}^{(1,1)}-
v_{-,s}^{(i,m)}G_{n,s}^{(1,2)}+
v_{-,s}^{(i+1,m)}G_{n,s}^{(2,1)}-
v_{+,s}^{(i-1,m)}G_{n,s}^{(2,2)} \right]=
$$
\begin{equation}
=f_{n}^{(1,1)}\delta_{i+1,m} \Delta_{n,0}-
f_{n}^{(1,2)}\delta_{i,m} \Delta_{n,0}, \label{qw17}
\end{equation}
$$
v_{-,n}^{(i,m)}+\sum_{s} \left[
v_{-,s}^{(i,m)}G_{n,s}^{(1,1)}-
v_{+,s}^{(i,m)}G_{n,s}^{(1,2)}+
v_{+,s}^{(i-1,m)}G_{n,s}^{(2,1)}-
v_{-,s}^{(i+1,m)}G_{n,s}^{(2,2)} \right]=
$$
\begin{equation}
=f_{n}^{(1,1)}\delta_{i,m} \Delta_{n,0}-
f_{n}^{(1,2)}\delta_{i+1,m} \Delta_{n,0}, \label{qw18}
\end{equation}
where
\begin{equation}
G_{n,s}^{(i,j)}=f_{n}^{(1,j)} {\cal T}_{n,s}^{(i)}-
F_{n}^{(1,j)} {\cal L}_{n,s}^{(i)}, \label{qw19}
\end{equation}

\begin{equation}
{\cal L}_{n,s}^{(1)}=\alpha \beta {\sum_{k,l}}^\prime
\gamma_{l} \sigma_{s,l}\sigma_{n,l}\Delta_{n,k} B_{k}/\epsilon_{k}=
\delta_{n,s} \frac{\sinh[(d-d_{\ast})\mu_{n}/a]}
{\sinh(d \mu_{n}/a)}, \;          \label{qw211}
\end{equation}
\begin{equation}
{\cal L}_{n,s}^{(2)}=\alpha \beta {\sum_{k,l}}^\prime (-1)^k
\gamma_{l} \sigma_{s,l}\sigma_{n,l}\Delta_{n,k} B_{k}/\epsilon_{k}=
=-\delta_{n,s} \frac{\sinh(d_{\ast}\mu_{n}/a)}
{\sinh(d \mu_{n}/a)}, \label{qw21}
\end{equation}
$$
{\cal T}_{n,s}^{(1)}=\alpha {\sum_{k,l}}^\prime
\gamma_{l} \sigma_{s,l} \Delta_{n,k} /
\left( \epsilon_{k} Z_{k,l} \right),
$$
$$
{\cal T}_{n,s}^{(2)}=\alpha {\sum_{k,l}}^\prime (-1)^k
\gamma_{l} \sigma_{s,l} \Delta_{n,k} /
\left( \epsilon_{k} Z_{k,l} \right),
$$
$$
{\cal T}_{n,s}^{(i)}=
\pi \frac {a} {b} \sum_{l=1}^{\infty}
\frac{ \theta_l^3 J_0^2(\theta_l) E_l^{(i)} \left(a/d,\nu_l \right)}
{\chi_l
\left(\lambda_n^2-\theta_l^{2}\right)
\left(\lambda_s^2-\theta_l^{2}\right)}
-\frac {1}{2} \delta_{n,s} E_2^{(i)}\left(a/d,\mu_n\right)+
$$
\begin{equation}
+\frac{ \pi a^2 }{\mu_n^2 b d}
\frac{\theta_1^{3} J_0^2\left(\theta_1\right)}
{\left(\lambda_n^2-\theta_1^{2}\right)
\left(\lambda_s^2-\theta_1^{2}\right)},  \label{qw21}
\end{equation}
$$
E_l^{(1)}(x,y)=\left\{
\begin{array}{lr}
\coth(y/x)/y-x/y^2,& \ \l=1, \\
\coth(y/x)/y,& \ l \neq 1,
\end{array}
\right.
$$
$$
E_l^{(2)}(x,y)=\left\{
\begin{array}{lr}
\sinh^{-1}(y/x)/y-x/y^2,& \ \l=1, \\
\sinh^{-1}(y/x)/y,& \ l \neq 1,
\end{array}
\right.
$$
$$
\theta_l=a\lambda_l/b, \;
\chi_l=\pi\lambda_l J_1^2(\lambda_l)/2,\; \nu_l=\sqrt{\theta_l^{2}-
\Omega^2}, \; \mu_l=\sqrt{\lambda_l^2-\Omega^2}.
$$

By comparing Eq.(\ref{qw7}) and Eq.(\ref{qw16}) one can deduce that the
electric field tangential components in the circular regions, through which
$i$-th cavity is connected with others elements of the system under
consideration , are only determined  via the fundamental mode ($E_{010}$)
amplitudes of all cavities.  The coefficients $v_{+,s}^{(i-1,m)}$,
determining the fields on the left side wall of the $i$-th cavity (the right
hole cross-section of the $i-1$-th disk),  and $v_{-,s}^{(i,m)}$,
determining the fields on the right side wall of the $i$-th cavity (the left
hole cross-section of the $i$-th disk), are proportional to the expansion
coefficients of tangential electric field with the complete set of functions
$\left\{J_1 \left(\lambda_s r/a\right)\right\}$.

Thus, the problem of coupled cavities has been rigorously reduced to the
problem of the coupling of electric fields (see, Eqs.(\ref{qw17},\ref{qw18})),
which are determined in circular regions
$r\leq a, \; z=D\times i, \;z=d+D\times i, \; i=-\infty,+\infty$.
As Eqs.(\ref{qw17},\ref{qw18}) indicate, immediately coupled are only four
fields: on the right and left cross-sections of the $i$-th disk, on the
left cross-section of the ($i+1$)-th disk and on the right cross-section of
the ($i-1$)-th disk.
This is a consequence of the fact that the $i$-th cavity in the chain is in
immediate contact with only two adjacent cavities ($i+1$ and $i-1$).
It can be deduced from Eqs.(\ref{qw17},\ref{qw18}) that the coefficients
$v_{+,s}^{(i,m)}$, $v_{-,s}^{(i,m)}$ obey the following symmetry conditions:
\begin{equation}
v_{\pm,s}^{(i,m)}=v_{\pm,s}^{(i+k,m+k)}, \label{qw22}
\end{equation}
\begin{equation}
v_{+,s}^{(m-1-k,m)}=v_{-,s}^{(m+k,m)}, \label{qw23}
\end{equation}
where $k$ is any integer. The correlation (\ref{qw22}) is a reflection of
translational symmetry of the system under consideration (field induced by the
$m$-th cavity on the $i$-th disk does not change during the simultaneous shift
of the cavity and disk numbers),
while the correlation (\ref{qw23})  is a reflection of considered mode
symmetry (field induced by the $m$-th cavity on the left cross-section of the
($m+k$)-th disk is equal to the field on the righ cross-section of the
($m-1-k$)-th disk).
Taking into account these correlations and introducing a new count of disks
to the right from the $i$-th cavity ($k=1,2,3 \ldots \infty$), we obtain
$$
\left(\omega_{0,1}^{2}-\omega^2\right) a_{0,1}^{(i)}=-
\omega_{0,1}^{2} \frac{2}{3\pi} \frac{a^3}{J_1^2(\lambda_1)b^2 d}\times
$$
\begin{equation}
\times
\left[
2 a_{0,1}^{(i)}\Lambda_{-,1}+
\sum_{k=1}^{\infty}
\left(a_{0,1}^{(i+k)}+a_{0,1}^{(i-k)} \right)
\left(\Lambda_{-,k+1}-\Lambda_{+,k}  \right)
\right], \label{qw24}
\end{equation}
where the normalized coupling coefficients are determined by the formulae:
\begin{equation}
\Lambda_{\pm,k}=J_0^2 \left(\lambda_1 a/b \right)
\sum_{s^{\prime}} w_{\pm,s^{\prime}}^{(k)}/
\left[
\lambda_{s^{\prime}}^2-{\left(\lambda_1 a/b \right)}^2
\right],\label{qw25}
\end{equation}
while field coupling over the semi-infinite disk number are described by
($w_{\pm,s}^{(k)}=\mp v_{\pm,s}^{(i+k-1,i)}=
\mp v_{\pm,s}^{(k-1,0)}$)
$$
w_{+,n}^{(1)}+\sum_{s} \left[
w_{+,s}^{(1)}G_{n,s}^{(1,1)}+
w_{-,s}^{(1)}G_{n,s}^{(1,2)}+
w_{-,s}^{(1)}G_{n,s}^{(2,2)} \right]=
$$
\begin{equation}
=\sum_{s} w_{-,s}^{(2)}G_{n,s}^{(2,1)}+
f_{n}^{(1,2)} 3\pi/
\left[
\lambda_n^2-{\left(\lambda_1 a/b \right)}^2
\right],\label{qw26}
\end{equation}
$$
w_{-,n}^{(1)}+\sum_{s} \left[
w_{-,s}^{(1)}G_{n,s}^{(1,1)}+
w_{+,s}^{(1)}G_{n,s}^{(1,2)}+
w_{-,s}^{(1)}G_{n,s}^{(2,1)} \right]=
$$
\begin{equation}
=\sum_{s} w_{-,s}^{(2)}G_{n,s}^{(2,2)}+
f_{n}^{(1,1)}   3\pi/
\left[
\lambda_n^2-{\left(\lambda_1 a/b \right)}^2
\right],\label{qw27}
\end{equation}
\begin{equation}
w_{+,n}^{(k)}+\sum_{s} \left[
w_{+,s}^{(k)}G_{n,s}^{(1,1)}+
w_{-,s}^{(k)}G_{n,s}^{(1,2)}\right]=
\sum_{s} \left[
w_{-,s}^{(k+1)}G_{n,s}^{(2,1)}+
w_{+,s}^{(k-1)}G_{n,s}^{(2,2)} \right], \label{qw28}
\end{equation}
\begin{equation}
w_{-,n}^{(k)}+\sum_{s} \left[
w_{-,s}^{(k)}G_{n,s}^{(1,1)}+
w_{+,s}^{(k)}G_{n,s}^{(1,2)}\right]=
\sum_{s} \left[
w_{+,s}^{(k-1)}G_{n,s}^{(2,1)}+
w_{-,s}^{(k+1)}G_{n,s}^{(2,2)} \right]. \label{qw29}
\end{equation}

In Eqs.(\ref{qw28},\ref{qw29}) $k=2,3 \ldots \infty$.

The closed set of equations (\ref{qw24}-\ref{qw29}) describes rigorously
the electrodynamic system under consideration. Eqs.(\ref{qw24})
describe the coupling of the infinite chain of the resonant circuits,
with the coupling coefficients  $\Lambda_{\pm,k}$ being frequency functions.

\section{Results of the Analysis and Numerical Simulation}

As mentioned above, the expansion into the Fourier series with the
short-circuit resonant cavity modes is complete for total electric field
within the cavity, but it is incapable of representing tangential electric
field on the hole, that is why $d_{\ast}$ has to satisfy the condition
$d_{\ast} < d$. Yet, as analysis and computer simulation indicate, in the
final formulae we may assume that $d_{\ast}=d$. It is connected with the fact,
that the value ${\cal L}_{n,s}^{(2)}$ as function $d_{\ast}$ has
a discontinuity, being equal to zero at $d_{\ast}=d$ and to finite value
at $d_{\ast}\ne d$.
The expression, determining ${\cal L}_{n,s}^{(2)}$ (see(\ref{qw21})), at
$d_{\ast} < d$  has a finite limit in the case $d_{\ast} \rightarrow d$:
$\lim_{d_{\ast} \rightarrow d}{\cal L}_{n,s}^{(2)}=-\delta_{n,s}$,
using which gives the correct results even at $d_{\ast}=d$.
The numerical simulations indicate that at  $10^{-7} < d_{\ast}\le d$
the calculation results depend on the value  $d_{\ast}$ only in the seventh
digit. In the case  $d_{\ast}=d$ --- ${\cal L}_{n,s}^{(1)}=0$ and, if we
put $G_{n,s}^{(2,i)}=0$ $(i=1,2)$ the Eqs.(\ref{qw26},\ref{qw27}) will
describe the field coupling on the disk dividing two equal cylindrical
cavities (see \cite{r8}-\cite{r10}).
Thus, the coupling of fields on various disks are described in
Eqs.(\ref{qw26},\ref{qw29}) by the terms which contain factors
$G_{n,s}^{(2,i)} \; (i=1,2)$.  This is confirmed as well by the fact that at
$a\rightarrow 0$ and $t=0$  $G_{n,s}^{(2,i)}\rightarrow 0 \; (i=1,2)$, while
$G_{n,s}^{(1,i)} \; (i=1,2)$ tend to constant values, independent of  $a$:
\begin{equation}
G_{n,s}^{(1,i)}=\lambda_{n} \int_0^\infty
d\theta
\frac{\theta^2 J_0^2(\theta)}
{\left(\lambda_n^2-\theta^2\right)
\left(\lambda_s^2-\theta^2\right)} -\frac{1}{2}\delta_{n,s}, \; (i=1,2).
\label{qw30}
\end{equation}

In this case Eqs.(\ref{qw26},\ref{qw29}) have the following solution
$$
w_{+,n}^{(1)}=w_{-,n}^{(1)}=6 \left(
\sin(\lambda_n)-\lambda_n\cos(\lambda_n)
\right)/\lambda_n^2,
$$
$$
w_{+,n}^{(k)}=w_{-,n}^{(k)}=0, \; k=2,3 \ldots \infty,
$$
at which $\Lambda_{-,1}=\Lambda_{+,1}=1$ and all the remaining values of the
normalized coupling coefficients are equal to zero. At these conditions the
set (\ref{qw24}) coincides with the equations describing an infinite cavity
chain obtained in the quasi-static approximations \cite{r14}-\cite{r16}.

The structure of Eqs.(\ref{qw26},\ref{qw27}), determining fields on the first
disk is different from that (\ref{qw28},\ref{qw29}) which determine fields on
the subsequent disks not only in the presence of the "driving" force, but also
in the nature of coupling. Fields on the first disk couple only with fields
on the second disk while fields on the disk, beginning from the second one,
couple with fields of two adjacent disks. This fact is associated with
transformation of infinite set of equations into semi-infinite one in
accordance with the field symmetry. Indeed, the coupling of the first disk
fields with the fields on the right cross-section of $(-1)$-th disk  in
accordance with the field symmetry has transformed into a "self-coupling",
described by the terms  $w_{-,s}^{(1)}G_{n,s}^{(2,2)}$ in (\ref{qw26}) and
$w_{-,s}^{(1)}G_{n,s}^{(2,1)}$ in (\ref{qw27}). This taken into account,
from the set
(\ref{qw24},\ref{qw29})) at $t \rightarrow \infty$ one can obtain the
equations describing the characteristics of a single cavity with two drift
tubes. In this case, owing to the field coupling, the frequency shift with
two tube will not be equal to the doubled frequency shift with one tube.

Eqs.(\ref{qw24}) have the solution of the kind
$a_{0,1}^{(n)}=a_{0} \exp(in\phi)$, where $a_{0}$ is the constant, while
$\phi$ is determined from the following equation:
$$
\left(\omega_{0,1}^{2}-\omega^2\right) =-
\omega_{0,1}^{2} \frac{4}{3\pi} \frac{a^3}{J_1^2(\lambda_1)b^2 d}\times
$$
\begin{equation}
\times \left[
\rho_0(\omega)+
\sum_{k=1}^{\infty} \rho_k(\omega)
\cos(k\phi)
\right], \label{qw32}
\end{equation}
where
$$\rho_0(\omega)=\Lambda_{-,1}(\omega),\; \rho_k (\omega)=
\left(\Lambda_{-,k+1}(\omega)-\Lambda_{+,k}(\omega)  \right).$$

From Eq.(\ref{qw32}) it follows that in the general case  in order to
determine the phase shift between cavities it is necessary that couplings
of all disks be taken into account. However, as numerical simulations
indicate, the contribution of "long range" couplings is small and one can
confine oneself to considering field couplings on the finite number of disks.
There, since we had used some symmetry relationships
((\ref{qw22})-(\ref{qw23})), it is necessary to observe the strict correlation
between the number of terms in the sum over  $k$ in Eq.(\ref{qw32}) and the
number of equations taken into account in ((\ref{qw26})-(\ref{qw29})).

Results of numerical analysis of Eq.(\ref{qw32}) are presented below.
Tab.\ref{tb1} gives the calculated values\footnote{Our results are in good
agreement both with the experimental data, given in \cite{r3}, and with
the calculation results performed within the program developed on the base
of partial region technique \cite{r17}} of phase shift (in degrees)
per cell for the cavity chains with such geometrical dimensions that
ensure phase shifts to occur close to $\phi_{0}=2\pi/3$ , $\pi/2$ ,
and $\pi/3$. The operation frequency is $f_0=2797.0$\,MHz
($\lambda_0=10.7183$\,cm). The column (1-0) presents the results of
calculations on the base of Eq.(\ref{qw32}) at $k=1$ in the case when fields
of various disks do not couple (in Eqs.(\ref{qw26},\ref{qw29})
$G_{n,s}^{(2,i)}=0 \; (i=1,2)$). The column (1) presents the results of
calculations at $k=1$ in the case when only "self-coupling" of the first
disk fields is taken into account in Eqs.(\ref{qw26},\ref{qw29}) (the 1-st
disk field couples with the field on the right cross-section of the (-1)-th
disk as described by the terms $w_{-,s}^{(1)}G_{n,s}^{(2,2)}$ in (\ref{qw26})
and $w_{-,s}^{(1)}G_{n,s}^{(2,1)}$ in (\ref{qw27})).
The columns (2)-(4) present the results of calculations at $k=2,3,4$.
Considering our transformation of the
infinite equation string into the semi-infinite one, we have: the results in
column (1-0) correspond to the case of non-coupling disks, (1) --- two disks,
(2) --- four, (3) --- six, (4) --- eight disks are coupled.
\begin{table}
\caption{Calculated values of the phase shift ($^\circ$) per one cavity
for various cavity chains}
\label{tb1}
\begin{center}
\begin{tabular}{|c|c|c|c|c|c|} \hline
\multicolumn{1}{|c|}{$a/\lambda_0$}&
\multicolumn{1}{|c|}{(1-0)}&
\multicolumn{1}{|c|}{(1)}&
\multicolumn{1}{|c|}{(2)}&
\multicolumn{1}{|c|}{(3)}&
\multicolumn{1}{|c|}{(4)}   \\ \hline
\multicolumn{6}{|c|}{$D=\lambda_0/3$}  \\ \hline
0.08 & 120.1630 & 120.0249 & 120.0119 & 120.0119 & 120.0119 \\ \hline
0.14 & 120.5835 & 120.0710 & 120.0123 & 120.0126 & 120.0126 \\ \hline
\multicolumn{6}{|c|}{$D=\lambda_0/4$}  \\ \hline
0.08 & 88.6985 & 90.1831 & 90.0103 & 90.0118 & 90.0118 \\ \hline
0.11 & 88.4270 & 90.4579 & 89.9777 & 89.9859 & 89.9859 \\ \hline
0.14 & 88.3926 & 90.9189 & 89.9929 & 90.0174 & 90.0176 \\ \hline
\multicolumn{6}{|c|}{$D=\lambda_0/6$}  \\ \hline
0.08 & 54.8735 & 61.9938 & 60.0590 & 60.0650 & 60.0655 \\ \hline
0.11 & 55.8546 & 63.8105 & 60.0845 & 60.0782 & 60.0833 \\ \hline
0.14 & 57.4583 & 65.7765 & 60.1612 & 60.0813 & 60.1005 \\ \hline
\end{tabular}
\end{center}
\end{table}

From Tab.(\ref{tb1}) it follows that the influence of coupling of different
disk fields on the buildup of a certain phase shift depends both on the
spacing of the disks and on the hole dimensions. Thus, for instance, in the
case of disk-loaded structures operating in the $\phi_{0}=\pi/3$\,mode,
even at small values of the hole radius, it is necessary to take into account
field coupling of four disks, while at large one --- six disks. In the
case of disk-loaded structures operating in the $\phi_{0}=\pi/2$\,mode it is
necessary to take into account field coupling of four disks. In the case of
the most commonly used disk-loaded structures operating in the
$\phi_{0}=2 \pi/3$\,mode only coupling of fields of two disks should be taken
into account for a broad range of hole radii.

The dependence of corresponding coefficients on frequency, in general,
seriously influences on the electrodynamic characteristics of the system
under consideration. Tab.\ref{tb2} presents calculation results of the
relationship of phase shift ($^\circ$) per cavity versus frequency
(dispersion relation) for a homogeneous disk-loaded wavequide with
$D=\lambda_{0}/3$ and $a/\lambda_{0}=0.14$.
The column ($\omega$) corresponds to the case $\rho_i=\rho_i(\omega)$
($i=0,1$), column ($\omega=\omega_{010}$) --- $\rho_i=\rho_i(\omega_{010})$,
column ($\omega=0$) --- $\rho=\rho_i(0)$ (quasistatic case).
From the calculations it follows that $\rho_i$vs.$\omega$ relationship, even
within the passband (cf.results in columns ($\omega$) and
($\omega=\omega_{010}$)), exercises an influence upon phase shift.
\begin{table}
\caption{Phase shift ($^\circ$) vs frequency for a cavity chain with
$D=\lambda_{0}/3$ ¨ $a/\lambda_{0}=0.14$}
\label{tb2}
\begin{center}
\begin{tabular}{|c|c|c|c|} \hline
\multicolumn{1}{|c|}{$f,\,ƒƒæ$}&
\multicolumn{1}{|c|}{($\omega$)}&
\multicolumn{1}{|c|}{($\omega=\omega_{010}$)}&
\multicolumn{1}{|c|}{($\omega=0$)}  \\ \hline
2.727 & 25.15 & 25.23 & 27.59  \\ \hline
2.737 & 45.12 & 45.24 & 48.19  \\ \hline
2.747 & 59.63 & 59.80 & 63.50  \\ \hline
2.757 & 72.20 & 72.48 & 77.01  \\ \hline
2.767 & 84.01 & 84.35 & 89.84  \\ \hline
2.777 & 95.57 & 96.03 & 102.72  \\ \hline
2.787 & 107.39 & 108.01 & 116.35 \\ \hline
2.797 & 120.07 & 120.93 & 131.94  \\ \hline
2.807 & 134.73 & 136.02 & 153.34  \\ \hline
2.877 & 155.07 & 157.81 &    \\ \hline
\end{tabular}
\end{center}
\end{table}

From the results above, one can deduce that cavity chains of any geometry
(homogeneous and inhomogeneous) with $D\ge\lambda_0/3$ and
$a/\lambda_0 \le 0.14$ can be described very accurately by the coupled
circuit model, wherein each resonant circuit is coupled to two :
$$
\left(\omega_{0,1}^{(i)2}-\omega^2\right) a_{0,1}^{(i)}=-
$$
\begin{equation}
-\omega_{0,1}^{(i)2} \left[
a_{0,1}^{(i)}\Gamma^{(i)}(\omega)-
\left(a_{0,1}^{(i+1)}\Gamma_{+}^{(i)}(\omega)+
a_{0,1}^{(i-1)}\Gamma_{-}^{(i)}(\omega) \right)
\right], \label{qw33}
\end{equation}
where the coefficients $\Gamma^{(i)}, \Gamma_{+}^{(i)},
\Gamma_{-}^{(i)}$ for $i$-th cavity will be determined by two values of the
radii of coupling holes, through which this cavity is connected with adjacent
ones, geometrical dimensions of the ($i-1,i,i+1$)-th cavities and frequency.
The results of studies of inhomogeneous cavity chains on the base of
Eq.(\ref{qw33}) will be presented in a future paper.

\section{Conclusion}

In this paper on the base of a rigorous electrodynamic approach we have
developed a mathematical model of a cylindrical cavity chain with
electric coupling. This model combines the model of the equivalent coupled
circuit chain and an accurate description of the non-resonant field influence.
The above approach can be also used in the case of magnetic coupling.
In this case the problem of accurate description of the potential fields on
the holes and slots (see, for example, \cite{r18}) will be easier, because
within the frame of the partial cross-over regions method the subset of
irrotational modes is a part of the complete set of modes that one has to
use to expand fields with. This technique is easily transformed for the
case of inhomogeneous structures. Then, there is a possibility to control
rigorously the effects of "long-range" coupling of cavities.

To this day, the equivalent circuit model was an only approximate one at
large couplings. In this case, one had to determine the circuit chain
parameters from the measured dispersion curves of the passbands. The above
method imbues one with hope that this model can give
sufficiently accurate description of the characteristics of the coupled
cavity chain at large couplings.


\newpage
\begin{thebibliography}{99}
\bibitem{r1} R.M.~Bevensee. Electromagnetic Slow Wave Systems. John
Wiley\&Sons, Inc.,New York-London-Sydney, 1964.
\bibitem{r2} A.D.~Grigorjev, V.B.~Yunkevich. Cavities and RF Cavity Slow
Systems. Moscow, "Radio and communications", 1984.
\bibitem{r3} O.A.~Valdner, N.P.~Sobenin, I.S.~Zverev et al. Disk Loaded
Waveguides. Reference Book. Moscow, Energoatomizdat, 1991.
\bibitem{r4} M.A.~Allen, G.S.~Kino. IRE Trans. Microwave Theory Tech.
1960. V.MTT-8. P.362-372.
\bibitem{r5} T.~Nishikava, S.~Giordano, D.~Carter. Rev.Sci.Instr.
1966. V.37. N.5. P.652-661.
\bibitem{r6} D.E.~Nagle, E.A.~Knapp, B.C.~Knapp. Rev.Sci.Instr.
1967. V.39. N.11. P.1583-1587.
\bibitem{r7} J.M.~Paterson, R.D.~Ruth, C.~Adolphsen et al. SLAC-PUB-5928,
1992.
\bibitem{r8} K.L.F.~Bane, R.L.~Gluckstern. SLAC-PUB-5783, 1992.
\bibitem{r9} M.I.~Ayzatsky. On two-cavity coupling. Preprint NSC KPTI 95-8,
1995.
\bibitem{r10} M.I.~Ayzatsky. Proc.14th Workshop on Charged Particle
Accelerators. Protvino, 1994, vol.1, p.240.
\bibitem{r11} M.I.~Ayzatsky. ZhTF. 1996, vol.66, in publication.
\bibitem{r12} A.G.~Tragov. Collected Series:
Accelerators, Moscow, Gosatomizdat, 1962, N.12, p.174-184.
\bibitem{r13} I.G.~Prohoda, V.I.~Lozyanoi, V.M.~Onufrienko et al.
Electromagnetic wave propagation in inhomogeneous waveguide systems.
Dnepropetrovsk, Dnepropetrovsk State University Publishing House, 1977.
\bibitem{r14} H.A.~Bathe. Phys. Rev. 1944. V.66. N.7. P.163-182.
\bibitem{r15} V.V.~Vladimirsky. ZhTF, 1947, v.17, N.11, p.1277-1282.
\bibitem{r16} A.I.~Akhiezer, Ya.B.~Fainberg. UFN, 1951, v.44, N.3, p.321-368.
\bibitem{r17} V.I.~Naidenko, E.V.~Gooseva. Radiotech. and Electr. 1987, V.32.
N.8, p.1735-1757.
\bibitem{r18} W-H.~Cheng, A.V.~Fedotov, R.L.~Gluckstern. Phys.Rev.E, 1995,
V.E52, N3, p.3127-3142.
\end{thebibliography}
\end{document}